\documentclass[conference]{IEEEtran}
\IEEEoverridecommandlockouts
\usepackage{cite}
\usepackage{amsmath,amssymb,amsfonts}
\usepackage{graphicx}
\usepackage{textcomp}
\usepackage{xcolor}
\usepackage{afterpage}
\usepackage{caption}

\usepackage{graphicx} 
\usepackage{booktabs} 
\usepackage{cellspace} 
\usepackage{booktabs} 
\usepackage{array} 
\usepackage{xcolor} 
\usepackage{graphicx} 

\usepackage{algorithm}
\usepackage{algpseudocode}
\usepackage{algorithm}
\usepackage{float} 
\usepackage{algpseudocode}
\usepackage{subcaption}
\usepackage{graphicx}
\usepackage{array}
\usepackage{balance}

\usepackage{amsmath}
\usepackage{fancyhdr}

\usepackage{todonotes}

\usepackage{booktabs}
\usepackage{caption}

\usepackage{placeins}
\usepackage{xcolor} 

\usepackage{multirow} 
\usepackage{graphicx} 

\def\BibTeX{{\rm B\kern-.05em{\sc i\kern-.025em b}\kern-.08em
    T\kern-.1667em\lower.7ex\hbox{E}\kern-.125emX}}

\usepackage{fancyhdr}

\fancypagestyle{firstpage}{%
  \fancyhf{}%

  \fancyhead[C]{\footnotesize THIS PAPER HAS BEEN ACCEPTED AT \underline{IEEE GCAIoT 2025} CONFERENCE AND FINAL VERSION IS NOT PUBLISHED YET.}
  \fancyfoot[C]{%
    \parbox{0.95\textwidth}{\centering\footnotesize
    ©2025 IEEE. The final published paper is copyrighted by IEEE and should be cited as:
    Amr Akmal Abouelmagd, Amr E Hilal, ``Emerging Paradigms for Securing Federated Learning Systems,''
    \textit{IEEE Global Conference on Artificial Intelligence \& Internet of Things (IEEE GCAIoT)}, 2025.
    Personal use of this material may be permitted and permission from IEEE must be obtained for all other uses.}
  }
}

\begin{document}

\typeout{*** MAIN JOB: \jobname ***}

\title{Emerging Paradigms for Securing Federated Learning Systems\\
}

\author{\IEEEauthorblockN{Amr Akmal Abouelmagd}
\IEEEauthorblockA{\textit{Department of Computer Science} \\
\textit{Tennessee Technological University}\\
Cookeville, TN, US \\
aabouelma42@tntech.edu}
\and
\IEEEauthorblockN{Amr Hilal}
\IEEEauthorblockA{\textit{Department of Computer Science } \\
\textit{Tennessee Technological University}\\
Cookeville, TN, US \\
ahilal@tntech.edu}
}

\maketitle
\thispagestyle{firstpage} 

\begin{abstract}
Federated Learning (FL) facilitates collaborative model training while keeping raw data decentralized, making it a conduit for leveraging the power of IoT devices while maintaining privacy of the locally collected data. However, existing privacy-preserving techniques present notable hurdles. Methods such as Multi-Party Computation (MPC), Homomorphic Encryption (HE), and Differential Privacy (DP) often incur high computational costs and suffer from limited scalability. This survey examines emerging approaches that hold promise for enhancing both privacy and efficiency in FL, including Trusted Execution Environments (TEEs), Physical Unclonable Functions (PUFs), Quantum Computing (QC), Chaos-Based Encryption (CBE), Neuromorphic Computing (NC), and Swarm Intelligence (SI). For each paradigm, we assess its relevance to the FL pipeline, outlining its strengths, limitations, and practical considerations. We conclude by highlighting open challenges and prospective research avenues, offering a detailed roadmap for advancing secure and scalable FL systems.
\end{abstract}


\section{Introduction}
Federated Learning (FL) \cite{mcmahan2017communication} is a distributed machine learning approach that enables multiple clients, predominantly IoT devices \cite{fang2021privacy}, \cite{khan2021federated}, to collaboratively train a shared global model without exchanging raw data. This method maintains model performance while optimizing for specific use cases, allowing IoT devices to process and learn from local data in a secure and privacy-preserving manner. By enabling decentralized learning, it encourages broader client participation and ensures sensitive data remains protected, which is critical in IoT environments. However, as FL has matured, research has revealed several vulnerabilities in classical FL architectures. Malicious entities can exploit these weaknesses to access private information or compromise system integrity through attacks that degrade global model performance. To address these challenges, researchers have integrated FL with various privacy-preserving techniques, including Multi-Party Computation (MPC) \cite{fang2021privacy}, Homomorphic Encryption (HE) \cite{zhang2020batchcrypt}, and Differential Privacy (DP) \cite{9069945}. These integrations aim to prevent attacks and create more resilient, efficient, secure, and robust FL systems. Despite significant efforts to secure the FL process, current FL systems encounter notable challenges and limitations when integrated with these security approaches \cite{kanagavelu2022fed}, \cite{pejic2022effect}, \cite{pustozerova2023analysing}:
\begin{itemize}
\item High communication overhead and latency in MPC schemes.
\item Scalability constraints in HE applications.
\item Model accuracy reduction in FL systems using DP schemes.
\end{itemize}

This survey highlights several emerging paradigms recently introduced in Federated Learning (FL) to address the limitations of existing techniques and achieve more secure, performant, and robust FL systems.
In this survey paper, we present the following contributions:
\begin{itemize}
\item A concise analysis of current limitations in the most commonly used security and privacy schemes for FL systems
\item A review of emerging techniques that can enhance security and privacy in FL while providing optimized processing with reduced communication and computational overhead
\item A mapping of each paradigm to specific stages in the FL pipeline to identify how their capabilities can create more efficient FL systems
\item A compilation of practical applications where each paradigm has been successfully deployed
\end{itemize}


\begin{figure*}[htbp]
    \centering
    \includegraphics[height=0.44\textheight]{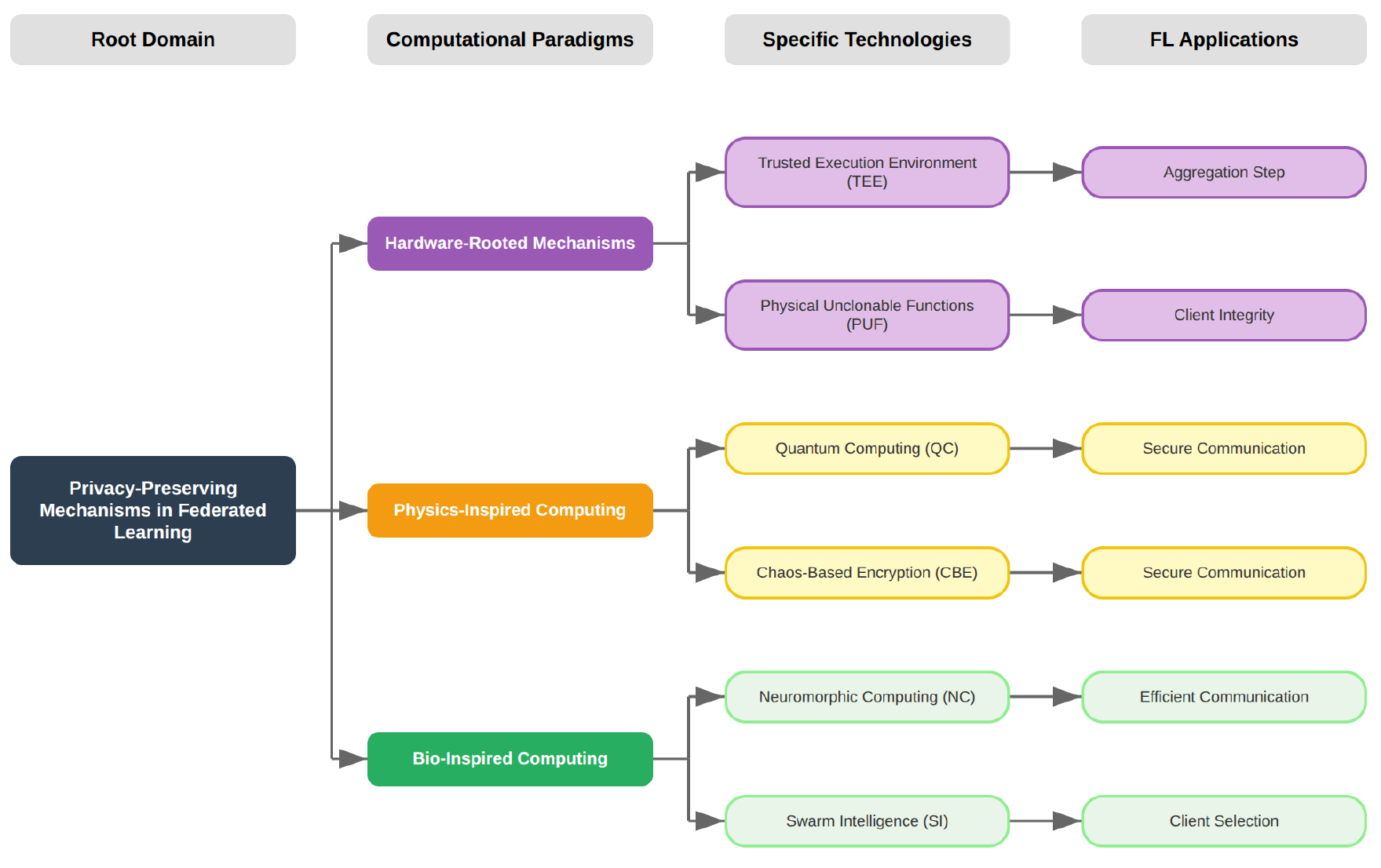}
    \caption{Conceptual diagram of the paper’s organization. The figure groups emerging paradigms in Federated Learning into distinct categories, lists the paradigms within each group, and maps them to the corresponding stages of the FL pipeline where they can be applied.}
    \label{fig:wide1}
\end{figure*}

The organization of this paper is illustrated in Figure \ref{fig:wide1}, which provides a high-level overview of the topics and emerging techniques discussed. We begin by examining the limitations of schemes used in current FL settings that hinder performance in Section II. In Section III, we review and explore emerging techniques and discuss their application in FL settings. Section IV addresses current challenges and open issues, while Section V presents possible future directions. Finally, Section VI concludes the paper.

\section{Why Go Beyond Traditional Privacy Approaches?}
While multiple techniques are widely used for achieving security and privacy in Federated Learning, they present notable limitations.
\subsubsection{Multi-Party Computation (MPC)}
Several challenges and limitations of the MPC scheme are discussed in \cite{liu2024survey}, \cite{truex2019hybrid}, where the authors identify issues such as substantial communication costs. This results from the high number of message exchanges required, making the approach inefficient and impractical across various Federated Learning (FL) settings—particularly when many clients are involved. Additionally, MPC introduces significant latency, which limits FL process performance by extending training completion times. Furthermore, these techniques do not protect against inference attacks that analyze function outputs to extract information about inputs.
In \cite{zhou2024secure}, the authors highlight the implementation difficulty of MPC, as it requires advanced technical expertise and deep understanding of cryptographic principles, creating barriers for many organizations.

\subsubsection{Homomorphic Encryption (HE)}
Homomorphic Encryption (HE) is a cryptographic technique that allows computations to be performed directly on encrypted data. The results, when decrypted, are identical to the outcome of performing the computations on the plaintext data itself, all without exposing the underlying information during processing.

Some drawback and limitations of HE have been cited and discussed in the literature \cite{manzoor2024survey}, \cite{Hu2024FederatedLearning}. These limitations include high computational cost incurred by HE, making it impractical for FL applications. Another limitation is that current HE schemes lack the efficiency and scalability required for deployment in practical FL settings. The authors mention that these limitations, combined with high computation overhead (observed in ShieldFL \cite{ma2022shieldfl}), make HE challenging for resource-constrained clients.
\subsubsection{Differential Privacy (DP)}
DP operates by adding calibrated noise to data or model updates to obscure individual contributions. In \cite{zhou2024secure}, the authors highlight several limitations of using Differential Privacy (DP) as a privacy enhancement mechanism in Federated Learning (FL). One significant limitation is that DP may compromise data utility due to the introduction of statistical noise; however, it is important to note that DP is not an encryption method. Additionally, DP does not offer the same cryptographic guarantees as Multi-Party Computation (MPC), which provides stronger protections for individual data during computation.

In \cite{el2022differential}, additional challenges associated with DP are discussed. The authors emphasize the inherent trade-off between privacy and accuracy: increasing noise levels improves privacy but can substantially degrade model performance. Conversely, insufficient noise may leave models vulnerable to privacy attacks. This concern is demonstrated in \cite{10.1145/3510032}, where the authors successfully reconstructed sensitive information from the original dataset by exploiting models trained with lower noise levels, illustrating the critical importance of selecting an appropriate privacy budget when applying DP.
Based on the limitations discussed for current privacy and security techniques used in FL, there is a clear need to explore new techniques and paradigms that enhance security and privacy in FL settings.
For each of the following paradigms, we explain the fundamental concept behind the technique in FL, discuss which FL stage or pipeline component the technique can address, examine the most recent research and development, and identify practical applications for the technique.

\section{Emerging Paradigms in Privacy-Preserving Federated Learning}
\subsection{Hardware-Rooted Mechanisms}
Several evolving technologies are built on specialized hardware designed to ensure security and privacy while increasing trust.
\subsubsection{Trusted Execution Environment (TEE)}
Trusted Computing \cite{Papa2011}, \cite{confidentialcomputing2022whitepaper} was developed to enable secure computing, safeguard user privacy, and ensure data protection. Initially, it relied on a dedicated hardware component designed to enhance platform security through a standardized interface. This component, known as the Trusted Platform Module (TPM) \cite{TPMSpecs}, enables a system to verify its own integrity and securely store cryptographic keys within a tamper-resistant hardware unit. However, TPM had several drawbacks, as it did not provide a separate isolated environment for execution by third parties, limiting its functionality to a predefined set of APIs.
Therefore, TEE was introduced in \cite{GlobalPlatformTEE2011} as a secure environment with memory and storage capabilities to achieve integrity-protected processing and execution \cite{sabt2015trusted}. In practice, TEEs have been widely deployed in the form of secure enclaves such as Intel SGX \cite{IntelSGX2016} and Arm TrustZone \cite{Ngabonziza2016TrustZone}, which are hardware-based systems that provide the security guarantees introduced by TEEs.

Several contributions leverage TEEs to achieve enhanced security and privacy in FL. In \cite{guan2024opsa}, the authors propose One-Pass Secure Aggregation (OPSA), a multi-round framework that achieves secure aggregation in a single pass, reducing communication overhead and improving FL process efficiency. Their approach also guarantees streamlined computation and verifiable aggregations. The work achieves a 2 to 10× speedup in multi-round aggregations with simultaneous result verification. They used TEEs to safely store and manage shared keys with clients, enabling the server to unmask and verify model updates in one step without requiring intensive computation inside the TEE, thus overcoming memory bottlenecks.
In \cite{zhu2025flsecure}, the authors explore techniques to address traditional FL challenges such as single points of failure in central servers and malicious attacks including Byzantine attacks. They propose FLSecure, a hybrid FL framework that integrates blockchain with TEEs to create a more reliable, secure, transparent, scalable, and decentralized FL system. By leveraging TEEs, FLSecure securely aggregates local model updates within isolated hardware environments, protecting them from unauthorized access and maintaining data integrity. Their approach develops a multi-TEE strategy that divides the global aggregation task into smaller subtasks distributed and executed simultaneously for faster processing and to address the memory constraints of TEEs.

To address the challenge of ensuring integrity in the FL training process, the authors in \cite{10773815} proposed a hardware-assisted federated learning framework to verify the integrity of client-side training in FL without exposing data or model updates. They leverage TEEs to enable task publishers to verify whether clients participating in the FL process have faithfully completed the training process. This prevents free-rider attacks, where dishonest clients upload model updates without performing local training, reduce local training rounds, or upload simpler models trained on less data to minimize effort while benefiting from the FL training process.
Based on the contributions discussed above, TEEs are best suited in the FL pipeline at the server side during the global model aggregation step, as they can prevent visibility into intermediate model updates. Since the aggregation step executes inside a secure environment with restricted access, this helps prevent malicious actors from tampering with the data. Even if the server or aggregator were compromised or acted with dishonest curiosity, TEEs ensure model updates are securely aggregated.
Furthermore, Intel SGX prevents multiple attacks such as replay attacks, which can occur in FL settings where the server attempts to roll back to an earlier, weaker global model state. Secure enclaves can detect this malicious behavior using trusted counters.

\subsubsection{Physical Unclonable Functions (PUFs)}
Physical Unclonable Functions \cite{herder2014physical} are hardware-based security primitives that leverage the inherent and unpredictable physical variations introduced during semiconductor manufacturing to generate unique identifiers or cryptographic keys. In essence, they serve as the hardware's fingerprint. They are unclonable due to the difficulty of replicating these variations, even for the original manufacturer. They are primarily used for secret key storage without requiring expensive hardware or large hardware components.
There are two types of PUFs: weak and strong PUFs. The main difference lies in the number of challenge-response pairs (CRP) the PUF can handle. Weak PUFs, such as SRAM PUFs, typically handle one challenge and are primarily used for secret key generation. Strong PUFs, such as Arbiter PUFs or Optical PUFs, can handle multiple CRPs and are primarily used for device authentication—for example, securely identifying devices to servers—without requiring additional hardware.

PUFs have been deployed in several applications. In \cite{vega2023resource}, the authors present a method to embed PUFs into the functional logic of FPGAs to reduce resource consumption. PUFs have also been utilized in IoT systems, where the authors in \cite{li2024practical} made IoT devices resistant to physical attacks such as side-channel attacks by leveraging PUFs, since no key material is stored in memory and CRPs are not stored in raw form.
Several contributions have leveraged PUFs to enhance security and privacy in FL systems. In \cite{aarella2023fortified}, the authors aimed to achieve faster and more efficient authentication processes in Collaborative Edge Computing, which shares many characteristics with FL. They used SRAM PUFs to generate unique, hardware-intrinsic digital fingerprints that serve as unique secret keys to create secure certificates for reliable authentication.

In \cite{gao2025ssl}, the authors aim to eliminate centralized key management systems required for authentication, particularly in resource-constrained environments. They present SSL-FL, a lightweight and secure authentication framework for Federated Learning. SSL-FL includes a component called SSL-PUF, which replaces centralized key management systems to reduce overhead during the authentication process. To achieve this, the central server and each edge client are equipped with twin SSL-PUF modules that are digitized and used for authentication.
Based on the contributions mentioned and the value that PUFs offer, they can be best utilized in early stages of the FL process, typically during the device authentication phase. They can verify client identity without storing keys, making them resistant to spoofing attacks.

\subsection{Physics-Inspired Computing}
\subsubsection{Quantum Computing (QC)}
Quantum computing \cite{steane1998quantum} is a paradigm of computation that harnesses quantum-mechanical phenomena to process information, offering the potential to solve certain problems more efficiently than classical computers. It is grounded in quantum mechanics \cite{fedak20091925}—a theoretical framework, first named by Max Born, that describes the motion and behavior of atomic and subatomic particles with the same level of generality and consistency that classical mechanics provides for macroscopic systems. In 2021, the authors of \cite{chehimi2022quantum} introduced the first quantum federated learning (QFL) framework, enabling decentralized training of quantum convolutional neural networks (QCNNs) on quantum data across multiple quantum clients. Their work marked the first practical integration of TensorFlow Federated with TensorFlow Quantum.
QM has been applied in Federated Learning settings for practical applications. In \cite{Liu2025PracticalQF}, the authors utilize QM to enhance privacy and scalability in the current quantum computing era. They propose a practical quantum federated learning framework designed for quantum networks. The framework leverages distributed quantum secret keys to secure local model updates and has been tested on a 4-client quantum network with a scalable structure.

In \cite{li2024privacy}, the authors address a critical challenge in classical FL which is data leakage attacks, particularly how gradient inversion attacks enable a central server to reconstruct clients' private data from shared gradients. They present innovative quantum communication protocols for federated learning that use quantum encoding and quantum communication instead of sending raw gradients, achieving a more secure FL process with efficient communication. They utilized several concepts from quantum mechanics, including the Quantum Bipartite Correlator \cite{li2024blind} and Quantum Communication Channels \cite{bennett2014quantum}.
Based on the contributions discussed, quantum mechanics-based approaches can enhance security and privacy in the communication steps executed during the FL process.

\subsubsection{Chaos-Based Encryption (CBE)}
Chaos-based encryption uses chaotic systems—systems that appear random but are deterministic and sensitive to initial conditions; small changes in input produce substantially different results. CBE uses the unpredictable, non-linear nature of chaotic maps \cite{yang2004survey} to secure data. One key advantage of chaos-based encryption techniques is their computational efficiency.
Although there is limited work in the area of CBEs with FL, there are some notable contributions worth discussing.
In \cite{arevalo2023chaotic}, the authors propose a secure FL technique that can handle incomplete and non-iid data by permitting encrypted sharing of missing feature distributions among participants. Furthermore, they add an additional layer of privacy for the FL setting that relies on chaotic map-based encryption techniques. They utilize the inherent complexity and unpredictability of chaos maps, which makes them resistant to conventional cryptographic attacks.
In the FL process, we can utilize CBEs to be resilient against inference attacks during transmission while exploiting the lightweight characteristics of CBEs. For example, during client-server communication steps, clients can encrypt their model updates before sending them to the server, and the central server can decrypt them, perform aggregation, and then share the newly aggregated results in a similar manner.

\subsection{Bio-inspired Computing}
\subsubsection{Neuromorphic Computing}
Neuromorphic Computing (NC) is an approach that mimics the structure and function of the human brain, primarily derived from biology and neuroscience. It uses Spiking Neural Networks (SNNs) \cite{maass1997networks} to model how neurons and synapses process information, enabling efficient computation similar to how the human brain works \cite{IBMNeuromorphic}.
SNNs have been deployed in practice. For example, Loihi \cite{8259423} is a neuromorphic chip developed by Intel, designed to mimic how the brain processes information. It is built on a 14-nanometer process and measures 60 mm². IBM developed TrueNorth \cite{OpenNeuromorphicTrueNorth} to emulate the brain, which consists of a 4096-core chip containing 1 million neurons and 256 million synapses. IBM researchers deployed deep neural network models to run on the TrueNorth architecture and achieved comparable precision with lower energy consumption \cite{DigitalMinds2016TrueNorth}.

In \cite{electronics12183984}, the authors discuss several problems present in current FL systems, such as the lack of techniques that can provide high privacy with low computational overhead, which hinders their applicability across different use cases (e.g., edge computing devices). They propose a novel FL framework that utilizes spiking neuron models and differential privacy. Since spiking neurons require less energy compared to traditional neural networks, their main approach mimics the local forward propagation process in a discrete way, similar to the transmission of nerve signals in the human brain. With this design and based on their experiments, they were able to reduce the accuracy of membership and property inference attacks while maintaining low energy cost.
Based on the discussion above, the sparsity characteristic of spiking neurons (where only a small number of neurons are active) can improve the efficiency of FL systems, primarily during the client-server communication phase, as the sparse activations produced by spiking neurons can lead to more compact model updates during training rounds.

\subsubsection{Swarm Intelligence}
Swarm Intelligence (SI) refers to the collective intelligence that emerges when multiple decentralized agents (or nodes) collaborate and share knowledge. These agents coordinate their actions without relying on a central authority. SI leverages self-organization, peer-to-peer collaboration, and local decision-making to tackle complex problems more effectively than individual agents or centralized systems \cite{warnat2021swarm}. 

Recently, SI has been applied in Federated Learning (FL). In \cite{raza2023proof}, the authors addressed the challenge of achieving reliable consensus when locally fine-tuned models produce conflicting predictions. Conventional methods like Byzantine Fault Tolerance \cite{castro1999practical} rely on simple majority voting, ignore confidence levels, and fail in tie scenarios. To overcome this, they proposed Proof of Swarm (PoSw), a lightweight consensus mechanism inspired by swarm intelligence. Each edge device contributes its prediction and confidence score, enabling consensus through iterative refinement. Evaluated on ECG classification with the MIT-BIH arrhythmia dataset \cite{moody2001impact}, PoSw showed fast convergence, resolved ties, and outperformed both individual local models and standard global models. These results highlight its effectiveness and viability for healthcare applications.

In \cite{khan2024swarm} the authors tackled the challenge that FL, in Industrial IoT and smart city environments, often suffers from slow convergence, poor generalization, and sensitivity to hyperparameter choices. To address this, they integrated Particle Swarm Optimization (PSO) \cite{kennedy1995particle} into the FL workflow, using swarm intelligence principles to automatically tune critical hyperparameters of local learning models in a distributed manner. Their experiments demonstrated that the proposed PSO-enhanced FL achieved faster convergence and higher classification accuracy compared to baseline FL without optimization, highlighting its effectiveness in large-scale, heterogeneous IoT scenarios.

Swarm intelligence is now also being explored in the context of large language models within federated learning. The authors in \cite{rjoub2025hybrid} addressed the challenge of efficiently deploying multimodal Large-Language Models (LLMs) across edge–cloud FL environments, where communication costs, resource limitations, and data heterogeneity hinder performance. They proposed a hybrid approach that combines Particle Swarm Optimization for selecting optimal edge devices and Ant Colony Optimization \cite{dorigo2007ant}, \cite{blum2005ant} for routing model updates between edge and cloud. Evaluated on multimodal tasks, this method achieved higher accuracy, reduced communication overhead, and improved client participation compared to standard FL strategies.

\section{Challenges and Open Issues}
\begin{figure}[h]
    \centering
    \includegraphics[width=\columnwidth]{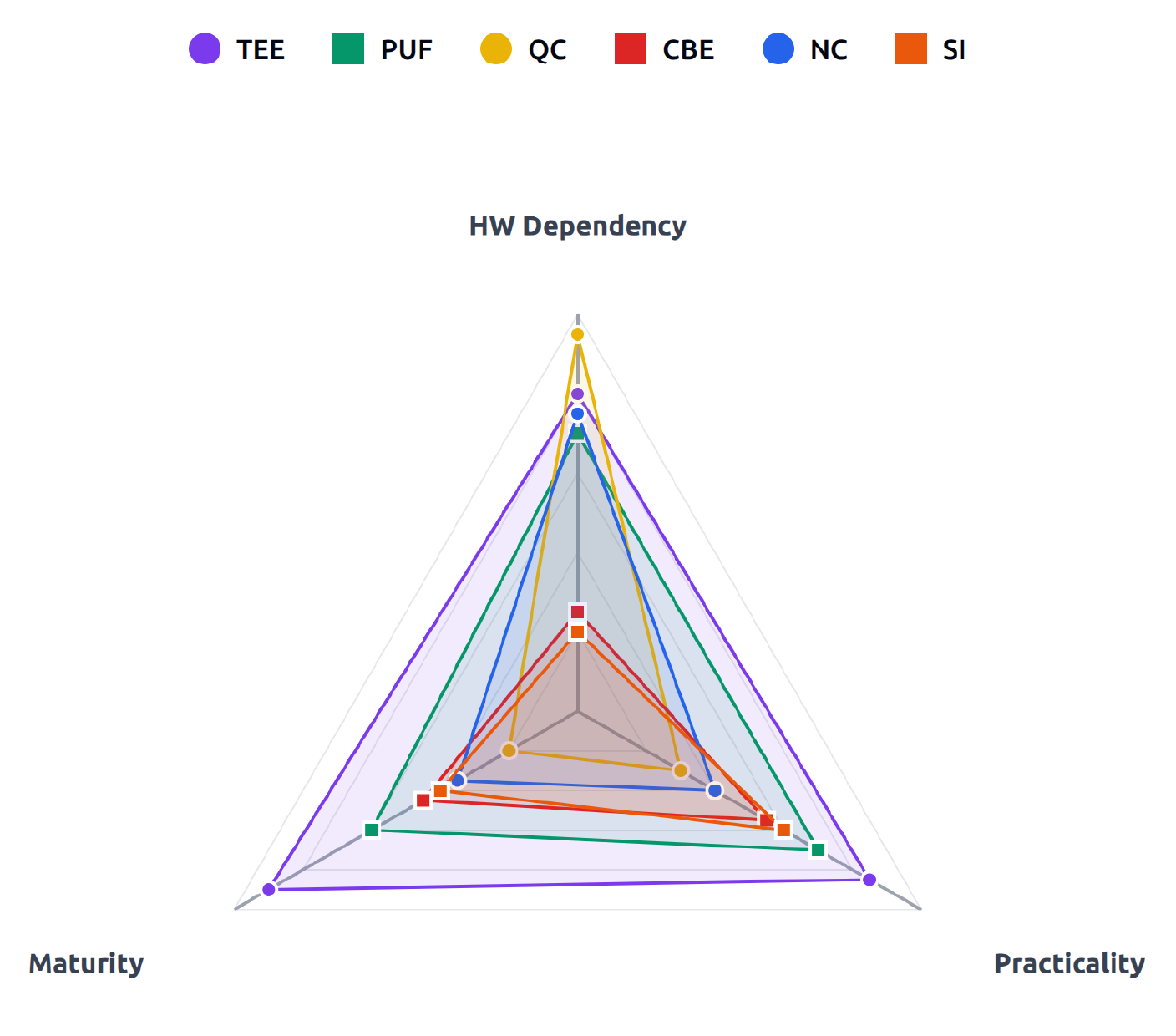}
    \caption{A visual comparison between the emerging privacy technologies in FL relative to each other with respect to hardware dependency, applicability, and maturity.}
    \label{fig:wide}
\end{figure}

While the advanced and emerging techniques discussed, illustrated in Figure \ref{fig:wide}, are promising and have shown potential for achieving more efficient FL systems, each paradigm presents specific challenges and limitations that require further investigation and exploration.
\subsection{Portability and Scalability Challenges}
Several advanced security and computing approaches face significant portability and scalability limitations that hinder their widespread adoption in federated learning systems. Trusted Execution Environments are constrained by limited enclave memory that restricts model complexity and dataset size, while their hardware dependency limits portability since not all clients can support TEEs in their systems. Addressing these challenges will help expand the use of TEEs across different clients, contributing to more robust and secure aggregation steps \cite{aublin2022towards}.
Similarly, most quantum computing systems currently available are difficult to scale and maintain efficient execution when thousands or millions of clients participate in FL systems. Their dependency on specific quantum hardware further limits portability for use across different clients \cite{article} \cite{gurung2023quantum}. Neuromorphic computing faces comparable constraints, being limited to specific hardware availability such as Loihi, which makes it impractical for use across distributed environments like FL settings with thousands of participating clients. Further research and investigation are required to enhance its applicability across various FL settings \cite{article2}.

\subsection{Reliability Constraints Due to Environmental Sensitivity}
Environmental sensitivity poses significant reliability constraints for certain security mechanisms in federated learning systems. Physical Unclonable Functions can be affected by the surrounding environment, which can impact their reliability. This represents an important issue that needs to be resolved to fully leverage the value of PUFs in building more secure FL systems \cite{Kim2025}. Similarly, chaos-based encryption's sensitivity to noise and precision errors can compromise its encryption reliability, which limits its usage in FL systems. CBEs require additional research and exploration to make them more applicable across different FL settings \cite{modelling6020025}.

Table~\ref{tab:paradigms_summary} summarizes key strength and limitations of each of the emerging technologies discussed.


\begin{table}[h]
\large
\centering
\caption{Summary of strengths and limitations of the emerging privacy technologies in Federate Learning}
\label{tab:paradigms_summary}
\resizebox{\columnwidth}{!}{%
\begin{tabular}{|p{2.5cm}|p{4.5cm}|p{5.5cm}|}
\hline
\textbf{Technology} & \textbf{Strengths} & \textbf{Limitations} \\ \hline
Trusted Execution Environments (TEEs) & Strong isolation and secure aggregation; prevents replay/tampering attacks & Limited enclave memory; hardware dependency; scalability challenges \\ \hline
Physical Unclonable Functions (PUFs) & Lightweight, unclonable hardware fingerprinting; strong device authentication without stored keys & Reliability affected by environmental factors; limited cryptographic versatility \\ \hline
Quantum Computing (QC) & Strong security from quantum properties; resistant to gradient inversion attacks; secure quantum communication & Requires specialized hardware; scalability is immature; noisy systems \\ \hline
Chaos-Based Encryption (CBE) & Lightweight and efficient; resilient to conventional cryptanalysis; good for transmission security & Sensitive to noise/precision; limited standardization and adoption in FL \\ \hline
Neuromorphic Computing (NC) & Energy-efficient processing; sparse activations reduce communication overhead; biologically inspired robustness & Hardware availability limited (Loihi, TrueNorth); lack of widespread integration in FL \\ \hline
Swarm Intelligence (SI) & Decentralized optimization for client selection and consensus; adaptive to heterogeneity and resilient to noisy/malicious clients & Higher computational cost for metaheuristics; convergence guarantees weaker than traditional optimization \\ \hline
\end{tabular}%
}
\end{table}

\section{Future Directions}
The integration of emerging paradigms into Federated Learning (FL), as shown in Figure~\ref{fig:wide1}, opens promising research avenues. However, several open problems and unexplored directions remain, which can shape the next generation of secure and efficient FL systems. Based on our analysis, we outline the following future directions:
\begin{itemize}
\item \textit{Hardware-Agnostic Security}: Many paradigms, such as QC and TEE, are limited by hardware dependencies. Research should aim to develop abstracted techniques that can reduce this hardware dependency, making them more generalizable and applicable across different settings.
\item \textit{Benchmarking and Standardization}: There is a need to establish standard benchmarks for evaluating the performance of emerging privacy-preserving techniques. These benchmarks must be secure, reliable, and efficient to help identify bottlenecks and direct attention toward resolving them. For example, paradigms such as SI require standardized evaluation to fairly assess their effectiveness in client selection, consensus, and robustness against adversarial behavior.
\item \textit{Hybrid Architectures}: Each emerging paradigm addresses privacy and efficiency from a different perspective. Innovative research that integrates and merges these different paradigms would be beneficial to maximize their utility and achieve efficient and secure FL systems.
\end{itemize}

\section{Conclusion}
Emerging paradigms such as TEEs, PUFs, QC, CBE, NC, and SI have broadened the landscape of privacy in Federated Learning by offering alternatives that promise stronger protections alongside scalability and efficiency. Their value lies not in replacing established methods but in expanding the toolkit available for different contexts. Yet, each approach carries its own technical and practical constraints—from hardware reliance to integration complexity—that must be carefully weighed before adoption. As summarized in Table~\ref{tab:paradigms_summary}, these paradigms are at different levels of maturity, highlighting both opportunities and limitations. For the field to progress, further work is needed to refine these techniques, validate them under real-world conditions, and clarify the trade-offs they entail. The challenge, and opportunity, is to translate their conceptual potential into reliable, deployable solutions.



\bibliographystyle{IEEEtran}   
\bibliography{references}  

\end{document}